\documentclass[aps,prd,preprint,showpacs,superscriptaddress]{revtex4}
\usepackage{graphicx}
\usepackage{bm}
\begin{document}

\title{Perturbation theory in Lagrangian hydrodynamics
for a cosmological fluid with velocity dispersion}

\author{Takayuki Tatekawa}
\email{tatekawa@gravity.phys.waseda.ac.jp}
\affiliation{Department of Physics, Waseda University,
3-4-1 Okubo, Shinjuku-ku, Tokyo 169-8555, Japan}
\author{Momoko Suda}
\affiliation{Department of Physics, Waseda University,
3-4-1 Okubo, Shinjuku-ku, Tokyo 169-8555, Japan}
\author{Kei-ichi Maeda}
\affiliation{Department of Physics, Waseda University,
3-4-1 Okubo, Shinjuku-ku, Tokyo 169-8555, Japan}
\affiliation{Advanced Research Institute for Science and Engineering,
Waseda University, 3-4-1 Okubo, Shinjuku-ku, Tokyo 169-8555, Japan}
\author{Masaaki Morita}
\affiliation{Theoretische Physik, Ludwig-Maximilians-Universit\"at,
Theresienstrasse 37, D-80333 M\"unchen, Germany}
\affiliation{Department of Physics, Ochanomizu University,
Otsuka, Bunkyo-ku, Tokyo 112-8610, Japan}
\affiliation{Advanced Research Institute for Science and Engineering,
Waseda University, 3-4-1 Okubo, Shinjuku-ku, Tokyo 169-8555, Japan}
\author{Hiroki Anzai}
\affiliation{NTT Communications, Yokosuka, Kanagawa, Japan}

\date{April 17, 2002}

\begin{abstract}
We extensively develop a perturbation theory for nonlinear
cosmological dynamics,
based on the Lagrangian description of hydrodynamics.
We solve hydrodynamic equations for a self-gravitating fluid
with pressure, given by a polytropic equation of state,
using a perturbation method up to second order.
This perturbative approach is an extension of the usual
Lagrangian perturbation theory for a pressureless fluid,
in view of inclusion of the pressure effect,
which should be taken into account
on the occurrence of velocity dispersion.
We obtain the first-order solutions in generic background universes
and the second-order solutions in wider range of a polytropic index,
whereas our previous work gives the first-order solutions
only in the Einstein-de Sitter background
and the second-order solutions for the polytropic index $4/3$.
Using the perturbation solutions,
we present illustrative examples of our formulation
in one- and two-dimensional systems,
and discuss how the evolution of inhomogeneities changes
for the variation of the polytropic index.
\end{abstract}

\pacs{04.25.Nx, 95.30.Lz, 98.65.Dx}

\maketitle

\section{Introduction}\label{sec:intro}

Hydrodynamics is a powerful tool to study various astrophysical
phenomena, from those associated with compact objects
up to large-scale structure formation.
 For example, when investigating gravitational instability
of cold dark matter for structure formation,
analyses are easier to handle by adopting a hydrodynamical description,
rather than by trying to solve the Boltzmann equation
of a self-gravitating N-particles system.
The linear perturbation theory of a homogeneous and isotropic
universe~\cite{weinberg,peebles,pad,coles,saco} is a typical case,
which gives a qualitative estimate for gravitational instability.
It is based on the Eulerian picture of hydrodynamics,
while approximations based on the Lagrangian hydrodynamics
has been recognized to be more useful,
such as the celebrated Zel'dovich approximation~\cite{zel,buchert92,munshi}.
This paper deals with an approximation theory of gravitational
instability based on the Lagrangian hydrodynamics.
Although the Zel'dovich approximation has been found to give an
accurate description up to the stage where density perturbations
grow to be unity,
it involves a serious shortcoming that it cannot be applied
after caustics in the density field are formed.
In the Zel'dovich approximation,
the fluid elements continue to move in the directions
that are determined by initial conditions all the time,
and consequently high density regions such as `pancakes'
cannot stay compact beyond the caustic formation,
while numerical simulations have shown the presence of clumps
with a very wide range in mass at any given time~\cite{davis}.
Moreover, once caustics in the density field are formed,
a hydrodynamical description itself is not valid in general.
Then, do we have to abandon a hydrodynamical description
and try to solve the Boltzmann equation,
or tackle N-body simulations?

In order to proceed with a hydrodynamical description
without the formation of caustics,
the `adhesion approximation'~\cite{gurbatov} has been proposed,
in which an artificial viscosity term is added
to the Zel'dovich approximation.
This modified approximation successfully describes the stage
where the original Zel'dovich approximation breaks down,
but the physical origin of the viscosity term should be clarified.
Some remarkable works have been done on this issue;
Buchert and Dom\'{\i}nguez~\cite{budo} argued,
by beginning with the collisionless Boltzmann equation~\cite{BT},
that the effect of velocity dispersion become important
beyond the caustics, where multi-stream flow arises,
and the presence of velocity dispersion of the flow can yield
pressure-like or viscosity terms;
Buchert et al.~\cite{bdp} showed how the viscosity term
is generated by a pressure-like force of a fluid
under the assumptions that the peculiar acceleration is
parallel to the peculiar velocity;
Dom\'{\i}nguez~\cite{domi00} clarified that a hydrodynamic
formulation is achieved via a spatial coarse graining
in a many-body gravitating system,
and the `adhesion approximation' can be derived by the expansion
of coarse-grained equations with respect to the smoothing length.
(See also Ref.~\cite{domi0106}.)
In these works, they also obtained implications for
an `equation of state', which is a phenomenological
relationship between kinematical pressure $P$ and
mass density $\rho$.
Buchert and Dom\'{\i}nguez~\cite{budo} found that,
if the effect of velocity dispersion is small and
the velocity dispersion is approximately isotropic,
the equation of state should take the form $P \propto \rho^{5/3}$;
Buchert et al.~\cite{bdp} showed that an adhesion-like equation
can be derived if the equation of state is assumed as
$P \propto \rho^2$;
Moreover, a plausible value of the polytropic index
$\gamma \equiv {\rm d} \ln P/{\rm d} \ln \rho$
has been found to be close to $5/3$ from
cosmological N-body simulations~\cite{domi0103}.

 From these aspects, it is of interest to extend
a Lagrangian perturbation scheme to a fluid with pressure,
and to explore how the scheme works as an approximation for
cosmological structure formation.
Actually, Adler and Buchert~\cite{adler}
and Morita and Tatekawa~\cite{moritate} have formulated
perturbation theory in the Lagrangian hydrodynamics,
taking into account the pressure effect under the assumption
that the pressure is a function of the mass density only.
In our earlier work~\cite{moritate},
imposing a polytropic relation as the equation of state,
we solved the Lagrangian perturbation equations up to second order
for cases where the equations are solved easily,
and showed illustrative examples with the solutions
in a one-dimensional system.
In particular, the second-order solutions were obtained
only for the case in which the polytropic index $\gamma$ is $4/3$,
while a plausible value of $\gamma$ seems to be larger,
as was mentioned above.

In this paper, we extend our earlier work
by solving the first-order perturbation equations
in generic background universes
and the second-order perturbation equations
for wider range of a polytropic index,
and by presenting illustrations in one- and two-dimensional systems.
We examine how the behavior of the perturbation solutions,
and the resultant evolution of inhomogeneities,
change for the variation of the polytropic index.
This enables us to discuss whether, or for what kind of
the equation of state, the adhesion-type approximation
is realized in the Lagrangian perturbation scheme.

This paper is organized as follows.
In Sec.~\ref{sec:basic}, we present Lagrangian hydrodynamic equations,
governing the system we consider.
In Sec.~\ref{sec:first}, the first-order perturbation equations are
derived and their solutions are shown,
not only in the Einstein-de Sitter background
but also in more generic backgrounds.
In Sec.~\ref{sec:second}, we obtain the second-order perturbation
equations and present their solutions in an approximate form
for $\gamma > 4/3$.
Section~\ref{sec:illust} provides illustrative examples
of our formulation in one- and two-dimensional models.
In Sec.~\ref{sec:discuss}, we discuss our results
and state our conclusions.

\section{Basic equations}\label{sec:basic}

In this section, we present hydrodynamic equations
in the Lagrangian description, which our approach stands on.
The matter model we consider is a self-gravitating fluid
with energy density $\rho$ and pressure $P$,
which arises in the presence of velocity dispersion.
Then the basic equations we start from are
\begin{equation}\label{contin}
\frac{\partial\rho}{\partial t}
+ 3\frac{\dot a}{a}\rho
+ \frac{1}{a}\nabla_{\bm{x}}
  \cdot (\rho \bm{v}) = 0 \,,
\end{equation}
\begin{equation}\label{euler}
\frac{\partial \bm{v}}{\partial t}
+ \frac{\dot a}{a}\bm{v}
+ \frac{1}{a} (\bm{v} \cdot
  \nabla_{\bm{x}}) \bm{v}
= \bm{g} - \frac{1}{\rho a}
  \nabla_{\bm{x}} P \,,
\end{equation}
\begin{equation}\label{rotg}
\nabla_{\bm{x}} \times \bm{g}
= 0 \,,
\end{equation}
\begin{equation}\label{poisson}
\nabla_{\bm{x}} \cdot \bm{g}
= -4\pi Ga (\rho - \rho_{\rm b}) \,,
\end{equation}
where $\bm{v}$ and $\bm{g}$ are the peculiar velocity
and the peculiar gravitational field, respectively,
which represent the deviation from a background,
homogeneous and isotropic universe.
The cosmic scale factor $a(t)$ and the energy density
$\rho_{\rm b}(t)$ of the background universe satisfy
the Friedmann equations
\begin{eqnarray}\label{friedmann1}
\left( \frac{\dot a}{a} \right)^2 &=& \frac{8\pi G}{3} \rho_{\rm b}
 - \frac{\mathcal{K}}{a^2} + \frac{\Lambda}{3} \,, \\
\label{friedmann2}
\frac{\ddot a}{a} &=& -\frac{4\pi G}{3} \rho_{\rm b}
 + \frac{\Lambda}{3} \,,
\end{eqnarray}
with a curvature constant $\mathcal{K}$
and a cosmological constant $\Lambda$.
In order to solve the hydrodynamic equations,
we must specify an equation of state.
Throughout this paper, we consider barotropic fluids,
in which the pressure $P$ is a function of the energy density only,
$P=P(\rho)$.

Introducing the Lagrangian time derivative
\[
\frac{\rm d}{{\rm d} t}
\equiv \frac{\partial}{\partial t} + \frac{1}{a}
 (\bm{v} \cdot \nabla_{\bm{x}}) \,,
\]
Eqs.~(\ref{contin}) and (\ref{euler}) become
\begin{equation}\label{continL}
\frac{{\rm d} \rho}{{\rm d} t}
+ 3\frac{\dot a}{a} \rho + \frac{\rho}{a}
 (\nabla_{\bm{x}} \cdot \bm{v}) =0 \,,
\end{equation}
\begin{equation}\label{eulerL}
\frac{{\rm d} \bm{v}}{{\rm d} t}
+ \frac{\dot a}{a} \bm{v}
= \bm{g}
- \frac{1}{\rho a} \nabla_{\bm{x}} P \,.
\end{equation}
In the Lagrangian hydrodynamics,
the coordinates $\bm{x}$ of the fluid elements are
represented in terms of Lagrangian coordinates $\bm{q}$ as
\begin{equation}
\bm{x} = \bm{q} + \bm{s} (\bm{q},t) \,,
\end{equation}
where $\bm{q}$ are defined as initial values of $\bm{x}$,
and $\bm{s}$ denotes the Lagrangian displacement vector
due to the presence of inhomogeneities.
The exact form of the energy density is then obtained from
Eq.~(\ref{continL}) as
\begin{equation}\label{exactrho}
\rho = \rho_{\rm b} J^{-1} \,,
\end{equation}
where $J \equiv \det (\partial x_i / \partial q_j)
= \det (\delta_{ij} + \partial s_i / \partial q_j)$
is the Jacobian of the coordinate transformation from
$\bm{x}$ to $\bm{q}$.
The peculiar velocity is
$\bm{v}=a \dot{\bm{s}}$,
and from Eq.~(\ref{eulerL}),
the peculiar gravitational field is written as
\begin{equation}
\bm{g} = a \left(
\ddot{\bm{s}}
+ 2\frac{\dot a}{a} \dot{\bm{s}}
- \frac{1}{a^2} \frac{{\rm d} P}{{\rm d} \rho} (\rho)
  \, J^{-1} \nabla_{\bm{x}} J \right) \,,
\end{equation}
where an overdot $(\dot{\mbox{ }})$ denotes ${\rm d}/{\rm d} t$.
Hence, from Eqs.~(\ref{rotg}) and (\ref{poisson}),
we obtain the following equations for $\bm{s}$:
\begin{equation}\label{rot-ddots}
\nabla_{\bm{x}} \times
\left( \ddot{\bm{s}}
+ 2\frac{\dot a}{a} \dot{\bm{s}}
\right) = 0 \,,
\end{equation}
\begin{equation}\label{div-ddots}
\nabla_{\bm{x}} \cdot \left(
\ddot{\bm{s}}
+ 2\frac{\dot a}{a} \dot{\bm{s}}
- \frac{1}{a^2} \frac{{\rm d} P}{{\rm d} \rho} (\rho)
  \, J^{-1} \nabla_{\bm{x}} J \right)
= -4\pi G\rho_{\rm b} (J^{-1} -1) \,.
\end{equation}
If we find solutions of Eqs.~(\ref{rot-ddots})
and (\ref{div-ddots}) for $\bm{s}$,
the dynamics of the system considered is completely determined.
Since these equations are highly nonlinear and hard to solve exactly,
we will advance a perturbative approach.
Remark that, in solving the equations for $\bm{s}$
in the Lagrangian coordinates $\bm{q}$,
the operator $\nabla_{\bm{x}}$ will be transformed
into $\nabla_{\bm{q}}$
by the following rule:
\begin{equation}
\frac{\partial}{\partial q_i} =
\frac{\partial x_j}{\partial q_i} \frac{\partial}{\partial x_j}
= \frac{\partial}{\partial x_i} +
\frac{\partial s_j}{\partial q_i} \frac{\partial}{\partial x_j} \,.
\end{equation}
%

\section{First-order solutions}\label{sec:first}

Hereafter we develop a perturbative approach for
the Lagrangian displacement vector $\bm{s}$
of the fluid elements.
In the first-order approximation,
Eqs.~(\ref{rot-ddots}) and (\ref{div-ddots}) become
\begin{equation}
\nabla_{\bm{q}} \times \left(
\ddot{\bm{s}}^{(1)} + 2\frac{\dot a}{a}
\dot{\bm{s}}^{(1)} \right)
= 0 \,,
\end{equation}
\begin{equation}
\nabla_{\bm{q}} \cdot \left(
\ddot{\bm{s}}^{(1)} + 2\frac{\dot a}{a}
\dot{\bm{s}}^{(1)}
- \frac{1}{a^2} \frac{{\rm d} P}{{\rm d} \rho} (\rho_{\rm b})
\,\nabla_{\bm{q}}
(\nabla_{\bm{q}} \cdot \bm{s}^{(1)})
\right)
= 4\pi G\rho_{\rm b} \nabla_{\bm{q}} \cdot
\bm{s}^{(1)} \,,
\end{equation}
where $\bm{s}^{(1)}$ denotes the first-order
displacement vector in the perturbative expansion.
Decomposing $\bm{s}^{(1)}$ into the longitudinal
and the transverse modes as
$\bm{s}^{(1)} = \nabla_{\bm{q}} S
+ \bm{S}^{\rm T}$ with
$\nabla_{\bm{q}} \cdot \bm{S}^{\rm T}=0$,
we have
\begin{equation}
\nabla_{\bm{q}} \times \left(
\ddot{\bm{S}^{\rm T}} + 2\frac{\dot a}{a}
\dot{\bm{S}^{\rm T}} \right)
= 0 \,,
\end{equation}
\begin{equation}
\nabla_{\bm{q}}^2 \left(
\ddot{S} + 2\frac{\dot a}{a} \dot{S}
- 4\pi G\rho_{\rm b} S
- \frac{1}{a^2} \frac{{\rm d} P}{{\rm d} \rho}(\rho_{\rm b})
\,\nabla_{\bm{q}}^2 S \right)
= 0 \,.
\end{equation}
These equations are reduced by imposing some adequate
boundary conditions to
\begin{equation}\label{ddotST}
\ddot{\bm{S}^{\rm T}} + 2\frac{\dot a}{a}
\dot{\bm{S}^{\rm T}} = 0 \,,
\end{equation}
\begin{equation}\label{ddotS}
\ddot{S} + 2\frac{\dot a}{a} \dot{S}
- 4\pi G\rho_{\rm b} S
- \frac{1}{a^2} \frac{{\rm d} P}{{\rm d} \rho}(\rho_{\rm b})
\,\nabla_{\bm{q}}^2 S = 0 \,.
\end{equation}
In our previous paper, we obtained the perturbation solutions
only for the Einstein-de Sitter background.
Here we solve the equations for the first-order perturbations
in generic background universes.
Equation~(\ref{ddotST}) can be integrated easily even in this case,
although an explicit form of the solutions is not presented here.
 For Eq.~(\ref{ddotS}), the Fourier transformation
with respect to the Lagrangian coordinates $\bm{q}$ yields
\begin{equation}\label{ddothatS}
\ddot{\widehat{S}} + 2\frac{\dot a}{a} \dot{\widehat{S}}
- 4\pi G\rho_{\rm b} \widehat{S}
+ \frac{1}{a^2} \frac{{\rm d} P}{{\rm d} \rho}(\rho_{\rm b})
\, |\bm{K}|^2 \widehat{S} = 0 \,,
\end{equation}
where $\widehat{(\cdot)}$ denotes the Fourier transform,
and $\bm{K}$ is a wavenumber associated with
the Lagrangian coordinates.
Replacing the time variable $t$ with $a$
and using the Friedmann equations~(\ref{friedmann1})
and (\ref{friedmann2}), we have
\begin{equation}
\left( \frac{8 \pi G \rho_{\rm b}}{3} a^2 - \mathcal{K}
       + \frac{\Lambda}{3} a^2 \right)
\frac{{\rm d}^2 \widehat{S}}{{\rm d} a^2}
+ \left(4 \pi G \rho_{\rm b} a - \frac{2 \mathcal{K}}{a} + \Lambda a \right)
\frac{{\rm d} \widehat{S}}{{\rm d} a}
+ \left( \frac{1}{a^2} \frac{{\rm d} P}{{\rm d} \rho} (\rho_{\rm b})
         \, |\bm{K}|^2 - 4 \pi G \rho_{\rm b} \right)
\widehat{S} = 0 \,.
\end{equation}
If we assume a polytropic equation of state $P=\kappa \rho^{\gamma}$
with a constant $\kappa$ and a polytropic index $\gamma$,
this equation becomes
\begin{equation}\label{diffa-hatS}
\left( \frac{2C_1}{a} - \mathcal{K} + \frac{\Lambda}{3} a^2 \right)
\frac{{\rm d}^2 \widehat{S}}{{\rm d} a^2}
+ \left( \frac{3C_1}{a^2} - \frac{2 \mathcal{K}}{a} + \Lambda a \right)
\frac{{\rm d} \widehat{S}}{{\rm d} a}
+ \left( \frac{C_2 |\bm{K}|^2}{a^{3\gamma -1}}
- \frac{3C_1}{a^3} \right) \widehat{S} = 0 \,,
\end{equation}
where $C_1 \equiv 4 \pi G \rho_{\rm b}(a_{\rm in})
\, a_{\rm in}^{\ 3} /3$
and $C_2 \equiv \kappa \gamma \rho_{\rm b}(a_{\rm in})^{\gamma-1}
\, a_{\rm in}^{\ 3(\gamma-1)}$.

Let us consider solving Eq.~(\ref{diffa-hatS}).
In the Einstein-de Sitter background,
where $\mathcal{K}=0$ and $\Lambda =0$,
the solutions of Eq.~(\ref{diffa-hatS}) are written
in a relatively simple manner.
They are, for $\gamma \ne 4/3$,
\begin{equation}\label{hatSbessel}
\widehat{S}(\bm{K},a) \propto a^{-1/4}
\, \mathcal{J}_{\pm 5/(8-6\gamma)}
\left( \sqrt{\frac{2C_2}{C_1}}
\frac{|\bm{K}|}{|4-3\gamma|}
\, a^{(4-3\gamma)/2} \right) \,,
\end{equation}
where $\mathcal{J}_{\nu}$ denotes the Bessel function of order $\nu$,
and for $\gamma=4/3$,
\begin{equation}\label{hatS43}
\widehat{S}(\bm{K},a) \propto
a^{-1/4 \pm \sqrt{25/16 - C_2 |\bm{K}|^2 / 2C_1}} \,.
\end{equation}
In the nonflat backgrounds with $\mathcal{K} \ne 0$ and $\Lambda =0$,
the solutions of Eq.~(\ref{diffa-hatS}) for
$\gamma=1$, $4/3$ can be written
in terms of Gauss' hypergeometric function $\mathcal{F}$ as
\begin{equation}
\widehat{S}(\bm{K},a) \propto a^{\beta} \, \mathcal{F}
\left( \alpha_1, \alpha_2, \alpha_3 ; \frac{\mathcal{K} a}{2 C_1}
\right) \,,
\end{equation}
where
\begin{eqnarray}
(\alpha_1, \alpha_2, \alpha_3, \beta) &=&
\left( -1+\sqrt{\frac{1}{4}
         +\frac{C_2 |\bm{K}|^2}{\mathcal{K}}} \,,
    \, -1-\sqrt{\frac{1}{4}
         +\frac{C_2 |\bm{K}|^2}{\mathcal{K}}} \,,
    \, -\frac{3}{2} \,, \, -\frac{3}{2}
\right) \,, \nonumber \\
 && \left( \frac{3}{2}+\sqrt{\frac{1}{4}
                      +\frac{C_2 |\bm{K}|^2}{\mathcal{K}}} \,,
        \, \frac{3}{2}-\sqrt{\frac{1}{4}
                      +\frac{C_2 |\bm{K}|^2}{\mathcal{K}}} \,,
        \, \frac{7}{2} \,, \, 1 \right)
 \quad \mbox{for } \gamma=1 \,,
\end{eqnarray}
\begin{eqnarray}
(\alpha_1, \alpha_2, \alpha_3, \beta) &=&
\left( \frac{3}{4} \pm \sqrt{\frac{25}{16}
                  -\frac{C_2 |\bm{K}|^2}{2C_1}} \,,
   \, -\frac{1}{4} \pm \sqrt{\frac{25}{16}
                  -\frac{C_2 |\bm{K}|^2}{2C_1}} \,,
\right. \nonumber \\
&& \left.
   \, 1 \pm \sqrt{\frac{25}{4}-\frac{2C_2 |\bm{K}|^2}{C_1}} \,,
   \, -\frac{1}{4} \pm \sqrt{\frac{25}{16}
                  -\frac{C_2 |\bm{K}|^2}{2C_1}}
\right)
 \quad \mbox{for } \gamma=4/3 \,.
\end{eqnarray}
In the flat ($\mathcal{K} = 0$) backgrounds with $\Lambda \ne 0$,
we can also write the solutions of Eq.~(\ref{diffa-hatS}) for
$\gamma=1/3$, $4/3$ in the form
\begin{equation}
\widehat{S}(\bm{K},a) \propto a^{\beta} \, \mathcal{F}
\left( \alpha_1, \alpha_2, \alpha_3 ; -\frac{\Lambda a^3}{6 C_1}
\right) \,,
\end{equation}
where
\begin{eqnarray}
(\alpha_1, \alpha_2, \alpha_3, \beta) &=&
\left( -\frac{1}{6}+\sqrt{\frac{1}{9}
                   -\frac{C_2 |\bm{K}|^2}{3\Lambda}} \,,
    \, -\frac{1}{6}-\sqrt{\frac{1}{9}
                   -\frac{C_2 |\bm{K}|^2}{3\Lambda}} \,,
    \, \frac{1}{6} \,, \, -\frac{3}{2}
\right) \,, \nonumber \\
 && \left( \frac{2}{3}+\sqrt{\frac{1}{9}
                      -\frac{C_2 |\bm{K}|^2}{3\Lambda}} \,,
        \, \frac{2}{3}-\sqrt{\frac{1}{9}
                      -\frac{C_2 |\bm{K}|^2}{3\Lambda}} \,,
        \, \frac{11}{6} \,, \, 1 \right)
 \quad \mbox{for } \gamma=1/3 \,,
\end{eqnarray}
\begin{eqnarray}
(\alpha_1, \alpha_2, \alpha_3, \beta) &=&
\left( \frac{7}{12} \pm \sqrt{\frac{25}{144}
                   -\frac{C_2 |\bm{K}|^2}{18C_1}} \,,
   \, -\frac{1}{12} \pm \sqrt{\frac{25}{144}
                   -\frac{C_2 |\bm{K}|^2}{18C_1}} \,,
\right. \nonumber \\
&& \left.
   \, 1 \pm \sqrt{\frac{25}{36}
      -\frac{2C_2 |\bm{K}|^2}{9C_1}} \,,
   \, -\frac{1}{4} \pm \sqrt{\frac{25}{16}
      -\frac{C_2 |\bm{K}|^2}{2C_1}}
\right)
 \quad \mbox{for } \gamma=4/3 \,.
\end{eqnarray}

Let us note the relation between the behavior of the above solutions
and the Jeans wavenumber, which is defined as
\[
K_{\rm J} \equiv \left(
\frac{4\pi G\rho_{\rm b} a^2}
     {{\rm d} P / {\rm d} \rho (\rho_{\rm b})} \right)^{1/2} \,.
\]
The Jeans wavenumber, which gives a criterion
whether a density perturbation with a wavenumber
will grow or decay with oscillation,
depends on time in general.
If the polytropic equation of state $P=\kappa\rho^{\gamma}$
is assumed,
\begin{equation}\label{kjeans}
K_{\rm J} = \sqrt{\frac{3C_1}{C_2}} \, a^{(3\gamma-4)/2} \,.
\end{equation}
Equation~(\ref{kjeans}) implies that, if $\gamma < 4/3$,
$K_{\rm J}$ will be infinitesimal and
density perturbations with any wavenumber will decay
in process of time,
and if $\gamma > 4/3$, all density perturbations will
grow to collapse.
This is confirmed by the form of the solutions,
Eq.~(\ref{hatSbessel}), by rewriting it as
\begin{equation}
\widehat{S}(\bm{K},a) \propto a^{-1/4}
\, \mathcal{J}_{\pm 5/(8-6\gamma)}
\left( \frac{\sqrt{6}}{|4-3\gamma|}
\frac{|\bm{K}|}{K_{\rm J}} \right) \,.
\end{equation}
However, this fact seems to be curious because
one may expect that, as the polytropic index $\gamma$ is larger,
the effect of the pressure would be stronger and consequently
the growth of density perturbations would be supressed
more effectively.
The unexpected result may be caused by construction of
the first-order approximation, in which the strength of
the pressure effect is determined only by the coefficient
$(1/a^2) \, {\rm d} P / {\rm d} \rho (\rho_{\rm b})$
in the fourth term of the left side of Eq.~(\ref{ddotS}).
The square of the `sound speed,' ${\rm d} P / {\rm d} \rho$,
which is contained in the coefficient,
is originally a function of $\rho$,
but now in the coefficient $\rho$ is replaced with $\rho_{\rm b}$
because of the first-order approximation.
Since $\rho_{\rm b} \propto a^{-3}$,
the coefficient decays sooner as the index $\gamma$ is larger,
and it leads to the consequence.
This problem may be resolved by trying higher-order approximations,
where the pressure effect is provided not only by the background density
but also by the presence of inhomogeneities.
Let us proceed to second order, noticing the above fact.

We should also note that the above curious behavior of
the perturbation solutions is seen in the Lagrangian coordinates,
not in the Eulerian coordinates.
In order to have a more precise discussion,
we have to transform the solutions into the form
in the Eulerian coordinates.
We will do so in a one-dimensional model in Sec.~\ref{sec:illust}.

\section{Second-order solutions}\label{sec:second}

In our previous paper, we derived the second-order solutions
only for the  case $\gamma=4/3$.
In this section, we obtain the second-order solutions for
the case $\gamma>4/3$ in an approximate form.
To second order, Eqs.~(\ref{rot-ddots}) and (\ref{div-ddots}) yield
\begin{equation}
\left[
\nabla_{\bm{q}} \times \left(
\ddot{\bm{s}}^{(2)} + 2\frac{\dot a}{a}
\dot{\bm{s}}^{(2)} \right)
\right]_i
= \epsilon_{ijk} s^{(1)}_{l\,,j} \left(
\ddot{s}^{(1)}_{k\,,l} + 2\frac{\dot a}{a} \dot{s}^{(1)}_{k\,,l}
\right) \,,
\end{equation}
\begin{eqnarray}
&& \ddot{s}^{(2)}_{i\,,i} + 2\frac{\dot a}{a} \dot{s}^{(2)}_{i\,,i}
- \frac{1}{a^2} \frac{{\rm d} P}{{\rm d} \rho} (\rho_{\rm b})
  \nabla_{\bm{q}}^2 s^{(2)}_{i\,,i}
- s^{(1)}_{j\,,i} \left(
\ddot{s}^{(1)}_{i\,,j} + 2\frac{\dot a}{a} \dot{s}^{(1)}_{i\,,j}
\right) \nonumber \\
&& + \frac{1}{a^2} \frac{{\rm d} P}{{\rm d} \rho} (\rho_{\rm b})
\left(
s^{(1)}_{i\,,ij} \nabla_{\bm{q}}^2 s^{(1)}_j
+ s^{(1)}_{i\,,jk} s^{(1)}_{j\,,ik}
+ s^{(1)}_{i\,,j} \nabla_{\bm{q}}^2 s^{(1)}_{j\,,i}
+ 2s^{(1)}_{i\,,j} s^{(1)}_{k\,,kij}
\right) \nonumber \\
&& + \frac{1}{a^2} \frac{{\rm d}^2 P}{{\rm d} \rho^2} (\rho_{\rm b})
  \rho_{\rm b} \left(
s^{(1)}_{i\,,i} \nabla_{\bm{q}}^2 s^{(1)}_{j\,,j}
+ s^{(1)}_{i\,,ik} s^{(1)}_{j\,,jk} \right)
= 4 \pi G \rho_{\rm b}
\left( s^{(2)}_{i\,,i} - \frac{1}{2}(s^{(1)}_{i\,,i})^2
       - \frac{1}{2} s^{(1)}_{i\,,j} s^{(1)}_{j\,,i}
\right) \,,
\end{eqnarray}
where $(\cdot)_{,i}$ denotes $\partial / \partial q_i$.
As in the first-order solutions,
we decompose $\bm{s}^{(2)}$ into the longitudinal
and the transverse modes as
$\bm{s}^{(2)} = \nabla_{\bm{q}} \zeta
+ \bm{\zeta}^{\rm T}$ with
$\nabla_{\bm{q}} \cdot \bm{\zeta}^{\rm T}=0$.
Then these equations are rewritten as
\begin{equation}\label{rot-ddotzetaT}
\left[
\nabla_{\bm{q}} \times \left(
\ddot{\bm{\zeta}^{\rm T}} + 2\frac{\dot a}{a}
\dot{\bm{\zeta}^{\rm T}} \right)
\right]_i
= \frac{1}{a^2} \frac{{\rm d} P}{{\rm d} \rho} (\rho_{\rm b})
\epsilon_{ijk} S_{,lj} \nabla_{\bm{q}}^2 S_{,kl} \,,
\end{equation}
\begin{eqnarray}\label{ddotzetaL}
&& \nabla_{\bm{q}}^2 \left(
\ddot{\zeta} + 2\frac{\dot a}{a} \dot{\zeta}
- 4\pi G\rho_{\rm b} \zeta
- \frac{1}{a^2} \frac{{\rm d} P}{{\rm d} \rho}(\rho_{\rm b})
\,\nabla_{\bm{q}}^2 \zeta \right) \nonumber \\
&& = 2 \pi G \rho_{\rm b} \left[ S_{,ij} S_{,ij}
   - (\nabla_{\bm{q}}^2 S)^2 \right]
- \frac{1}{a^2} \frac{{\rm d} P}{{\rm d} \rho} (\rho_{\rm b})
\left( \nabla_{\bm{q}}^2 S_{,i} \nabla_{\bm{q}}^2 S_{,i}
       + S_{,ijk} S_{,ijk} + 2S_{,ij} \nabla_{\bm{q}}^2 S_{,ij}
\right) \nonumber \\
&& \quad
- \frac{1}{a^2} \frac{{\rm d}^2 P}{{\rm d} \rho^2} (\rho_{\rm b})
  \rho_{\rm b} \left( \nabla_{\bm{q}}^2 S
                      \nabla_{\bm{q}}^2 \nabla_{\bm{q}}^2 S
  + \nabla_{\bm{q}}^2 S_{,i} \nabla_{\bm{q}}^2 S_{,i} \right) \,,
\end{eqnarray}
where we have neglected the first-order transverse perturbation
$\bm{S}^{\rm T}$ for simplicity,
and used Eq.~(\ref{ddotS}).
Taking the rotation of Eq.~(\ref{rot-ddotzetaT}), we obtain
\begin{equation}\label{lap-ddotzetaT}
-\nabla_{\bm{q}}^2 \left(
\ddot{\zeta^{\rm T}_i} + 2\frac{\dot a}{a}
\dot{\zeta^{\rm T}_i} \right)
= \frac{1}{a^2} \frac{{\rm d} P}{{\rm d} \rho} (\rho_{\rm b})
  \left( S_{,ijk} \nabla_{\bm{q}}^2 S_{,jk}
         + S_{,ij} \nabla_{\bm{q}}^2 \nabla_{\bm{q}}^2 S_{,j}
         - \nabla_{\bm{q}}^2 S_{,j} \nabla_{\bm{q}}^2 S_{,ij}
         - S_{,jk} \nabla_{\bm{q}}^2 S_{,ijk}
  \right) \,.
\end{equation}
The Fourier transform of Eqs.~(\ref{ddotzetaL}) and
(\ref{lap-ddotzetaT}) gives
\begin{eqnarray}
&& -|\bm{K}|^2
\left( \ddot{\widehat{\zeta}} + 2 \frac{\dot a}{a}
\dot{\widehat{\zeta}} - 4 \pi G \rho_{\rm b} \widehat{\zeta}
+ \frac{1}{a^2} \frac{{\rm d} P}{{\rm d} \rho} (\rho_{\rm b})
  \, |\bm{K}|^2 \widehat{\zeta}
\right) \nonumber \\
&& = \frac{1}{(2 \pi)^3} \int_{-\infty}^{\infty}
{\rm d}^3 \bm{K}'
\, \widehat{S} (\bm{K}',t) \, \widehat{S} (\bm{K}-\bm{K}', t)
\, \Biggl[ 2 \pi G \rho_{\rm b} \left\{
           \left( \bm{K}' \cdot (\bm{K}-\bm{K}') \right)^2
         - |\bm{K}'|^2 |\bm{K}-\bm{K}'|^2 \right\}
\nonumber \\
&& \quad
+ \frac{1}{a^2} \frac{{\rm d} P}{{\rm d} \rho} (\rho_{\rm b})
\left\{ |\bm{K}'|^2 |\bm{K}-\bm{K}'|^2
        \left( \bm{K}' \cdot (\bm{K} - \bm{K}') \right) +
        \left( \bm{K}' \cdot (\bm{K} - \bm{K}') \right)^3
\right. \nonumber \\
&& \qquad
\left.  + 2 |\bm{K} - \bm{K}'|^2
        \left( \bm{K}' \cdot (\bm{K} - \bm{K}') \right)^2
\right\} \nonumber \\
&& \quad
+ \frac{1}{a^2} \frac{{\rm d}^2 P}{{\rm d} \rho^2} (\rho_{\rm b})
\left\{ |\bm{K}'|^2 |\bm{K} - \bm{K}'|^4 +
        |\bm{K}'|^2 |\bm{K} - \bm{K}'|^2
        \left( \bm{K}' \cdot (\bm{K} - \bm{K}') \right)
\right\} \Biggr] \,,
\label{ddot-hatzetaL}
\end{eqnarray}
\begin{eqnarray}
|\bm{K}|^2 \left(
\ddot{\widehat{\zeta^{\rm T}_i}} + 2 \frac{\dot{a}}{a}
\dot{\widehat{\zeta^{\rm T}_i}} \right)
&=& - \frac{i}{(2 \pi)^3} \frac{1}{a^2}
\frac{{\rm d} P}{{\rm d} \rho} (\rho_{\rm b})
\int_{-\infty}^{\infty} {\rm d}^3 \bm{K}'
\,\widehat{S}(\bm{K}', t) \,\widehat{S}(\bm{K}-\bm{K}', t)
\nonumber \\
&& \cdot |\bm{K}-\bm{K}'|^2
   \left(\bm{K}' \cdot (\bm{K}-\bm{K}') \right)
   \biggl[ K'_i \left\{ \bm{K}' \cdot (\bm{K}-\bm{K}')
           + |\bm{K} - \bm{K}'|^2
   \right\} \nonumber \\
&& \quad - (K_i-K'_i) \left\{ \bm{K}' \cdot (\bm{K}-\bm{K}')
   + |\bm{K}'|^2 \right\} \biggr] \,.
\label{ddot-hatzetaT}
\end{eqnarray}
Using the Green functions $G(\bm{K},t,t')$
and $G^{\rm T}(t,t')$,
Eqs.~(\ref{ddot-hatzetaL}) and (\ref{ddot-hatzetaT})
are solved in the form
\begin{equation}
\widehat{\zeta}(\bm{K},t)
= -\frac{1}{|\bm{K}|^2}
\int^t {\rm d} t' \  G(\bm{K},t,t')
\  \widehat{Q} (\bm{K},t') \,,
\end{equation}
\begin{equation}
\widehat{\zeta^{\rm T}_i}(\bm{K},t)
= \frac{1}{|\bm{K}|^2}
\int^t {\rm d} t' \  G^{\rm T}(t,t')
\  \widehat{Q^{\rm T}_i} (\bm{K},t') \,,
\end{equation}
where $\widehat{Q} (\bm{K},t)$ and
$\widehat{Q^{\rm T}_i} (\bm{K},t)$
denote the right-hand side of Eqs.~(\ref{ddot-hatzetaL})
and (\ref{ddot-hatzetaT}), respectively.

In order to obtain an explicit form of the second-order solutions,
we assume the Einstein-de Sitter background with a normalization
so that $a(t) =t^{2/3}$,
and the equation of state as $P = \kappa \rho^{\gamma}$.
The first-order solutions are then
\begin{equation}
\bm{S}^{\rm T} \propto
\  {\rm const.} \,, \  t^{-1/3} \,,
\end{equation}
\begin{equation}\label{hatS-t}
\widehat{S}(\bm{K},t) \propto
\  t^{-1/6} \mathcal{J}_{\pm 5/(8-6\gamma)}
\left( A |\bm{K}| t^{-\gamma+4/3} \right)
\quad \mbox{for } \gamma \ne 4/3 \,,
\end{equation}
\begin{equation}\label{hatS43-t}
\widehat{S}(\bm{K},t) \propto
\  t^{-1/6 \pm \sqrt{25/36 - B |\bm{K}|^2}}
\quad \mbox{for } \gamma = 4/3 \,,
\end{equation}
where
\[
A \equiv \frac{1}{|4-3\gamma|} \sqrt{\frac{2C_2}{C_1}} \,,
\quad
B \equiv \frac{2C_2}{9C_1} \,.
\]
These first-order solutions yield the Green functions
in the following form:
\begin{equation}
G^{\rm T} (t,t') = 3 (t' - t^{-1/3} t'^{4/3}) \,,
\end{equation}
\begin{eqnarray}
G (\bm{K}, t, t') &=&
- \frac{\pi}{2 \sin \nu \pi} \left( -\gamma + \frac{4}{3} \right)^{-1}
  t^{-1/6} t'^{7/6} \biggl[
  \mathcal{J}_{-\nu} (A |\bm{K}| t^{-\gamma + 4/3})
  \mathcal{J}_{\nu} (A |\bm{K}| t'^{-\gamma + 4/3}) \nonumber \\
&& \qquad \qquad
 -\mathcal{J}_{\nu} (A |\bm{K}| t^{-\gamma + 4/3})
  \mathcal{J}_{-\nu} (A |\bm{K}| t'^{-\gamma + 4/3})
                    \biggr]  \quad \mbox{for } \gamma \ne 4/3 \,,
\end{eqnarray}
\begin{eqnarray}
G (\bm{K}, t, t') &=&
-\frac{1}{2} \left( \frac{25}{36} - B |\bm{K}|^2
             \right)^{-1/2} t^{-1/6} t'^{7/6} \biggl(
   t^{-\sqrt{25/36-B|\bm{K}|^2}}
   t'^{\sqrt{25/36-B|\bm{K}|^2}} \nonumber \\
&& \qquad \qquad
 - t^{\sqrt{25/36-B|\bm{K}|^2}}
   t'^{-\sqrt{25/36-B|\bm{K}|^2}}
\biggr) \quad \mbox{for } \gamma = 4/3 \,,
\end{eqnarray}
where we have assumed that $\nu \equiv 5/(8-6\gamma)$ is not an integer.
If we write the first-order solution as
$\widehat{S}(\bm{K},t)
= D^+(\bm{K},t) C^+(\bm{K}) + D^-(\bm{K},t) C^-(\bm{K})$,
where $D^{\pm}(\bm{K},t)$ are given
by the form of Eqs.~(\ref{hatS-t}) and (\ref{hatS43-t}),
we obtain
\begin{eqnarray}
\widehat{\zeta^{\rm T}_i}(\bm{K},t) &=&
-\frac{i}{(2\pi)^3} \frac{1}{|\bm{K}|^2}
\int^{\infty}_{-\infty} {\rm d}^3 \bm{K}'
\,E^{\rm T}(\bm{K},\bm{K}',t)
\Bigl(C^+(\bm{K}') C^+(\bm{K}-\bm{K}') \nonumber \\
&& \quad + C^+(\bm{K}') C^-(\bm{K}-\bm{K}')
         + C^-(\bm{K}') C^+(\bm{K}-\bm{K}')
         + C^-(\bm{K}') C^-(\bm{K}-\bm{K}')
\Bigr) \nonumber \\
&& \quad \cdot |\bm{K}-\bm{K}'|^2
\left( \bm{K}' \cdot(\bm{K}-\bm{K}') \right)
\biggl[ K'_i \left\{
             \bm{K}'\cdot(\bm{K}-\bm{K}')
             + \bm{K}' |\bm{K}-\bm{K}'|^2 \right\} \nonumber \\
&& \qquad - (K_i - K'_i) \left\{
          \bm{K}'\cdot(\bm{K}-\bm{K}') + |\bm{K}'|^2 \right\}
\biggr] \,,
\label{hatzetaT}
\end{eqnarray}
\begin{eqnarray}
\widehat{\zeta}(\bm{K},t) &=&
-\frac{1}{(2\pi)^3} \frac{1}{|\bm{K}|^2}
\int^{\infty}_{-\infty} {\rm d}^3 \bm{K}' \,
\Bigl(C^+(\bm{K}') C^+(\bm{K}-\bm{K}')
    + C^+(\bm{K}') C^-(\bm{K}-\bm{K}') \nonumber \\
&& \quad
    + C^-(\bm{K}') C^+(\bm{K}-\bm{K}')
    + C^-(\bm{K}') C^-(\bm{K}-\bm{K}')\Bigr) \nonumber \\
&& \quad \cdot
\biggl[ E(\bm{K},\bm{K}',t)
 \left\{ \left(\bm{K}' \cdot(\bm{K}-\bm{K}')\right)^2
 - |\bm{K}'|^2 |\bm{K}-\bm{K}'|^2 \right\} \nonumber \\
&& \quad
+ F_1(\bm{K},\bm{K}',t)
\left\{ |\bm{K}'|^2 |\bm{K} - \bm{K}'|^2
        \,\bm{K}'\cdot (\bm{K}-\bm{K}')
        + \left(\bm{K}' \cdot(\bm{K}-\bm{K}')\right)^3
\right. \nonumber \\
&& \qquad
\left.  + 2|\bm{K}-\bm{K}'|^2
 \left( \bm{K}'\cdot(\bm{K}-\bm{K}') \right)^2
\right\} \nonumber \\
&& \quad
+ F_2(\bm{K},\bm{K}',t)
\left\{ |\bm{K}'|^2 |\bm{K}-\bm{K}'|^4
      + |\bm{K}'|^2 |\bm{K}-\bm{K}'|^2
        \,\bm{K}'\cdot (\bm{K}-\bm{K}')
\right\}
\biggr] \,,
\label{hatzeta}
\end{eqnarray}
where the time-dependent factors are given as
\begin{eqnarray}
E^{\rm T}(\bm{K},\bm{K}',t) &=&
\int^t \frac{{\rm d} t'}{a^2(t')}
\frac{{\rm d} P}{{\rm d} \rho}(\rho_{\rm b}(t')) \,G^{\rm T}(t,t')
\,\Bigl(D^+(\bm{K}',t') D^+(\bm{K}-\bm{K}',t') \nonumber \\
&& \quad 
      + D^+(\bm{K}',t') D^-(\bm{K}-\bm{K}',t')
      + D^-(\bm{K}',t') D^+(\bm{K}-\bm{K}',t') \nonumber \\
&& \quad
      + D^-(\bm{K}',t') D^-(\bm{K}-\bm{K}',t') \Bigr) \,,
\label{ET(K,K',t)}
\end{eqnarray}
\begin{eqnarray}
E(\bm{K},\bm{K}',t) &=&
\int^t {\rm d} t' \, 2\pi G\rho_{\rm b}(t')
\,G(\bm{K},t,t')
\,\Bigl(D^+(\bm{K}',t') D^+(\bm{K}-\bm{K}',t') \nonumber \\
&& \quad
      + D^+(\bm{K}',t') D^-(\bm{K}-\bm{K}',t')
      + D^-(\bm{K}',t') D^+(\bm{K}-\bm{K}',t') \nonumber \\
&& \quad
      + D^-(\bm{K}',t') D^-(\bm{K}-\bm{K}',t') \Bigr) \,,
\end{eqnarray}
\begin{eqnarray}
F_1(\bm{K},\bm{K}',t) &=&
\int^t \frac{{\rm d} t'}{a^2(t')}
\frac{{\rm d} P}{{\rm d} \rho}(\rho_{\rm b}(t'))
\,G(\bm{K},t,t')
\,\Bigl(D^+(\bm{K}',t') D^+(\bm{K}-\bm{K}',t') \nonumber \\
&& \quad
      + D^+(\bm{K}',t') D^-(\bm{K}-\bm{K}',t')
      + D^-(\bm{K}',t') D^+(\bm{K}-\bm{K}',t') \nonumber \\
&& \quad
      + D^-(\bm{K}',t') D^-(\bm{K}-\bm{K}',t') \Bigr) \,,
\end{eqnarray}
\begin{eqnarray}
F_2(\bm{K},\bm{K}',t) &=&
\int^t \frac{{\rm d} t'}{a^2(t')}
\frac{{\rm d}^2 P}{{\rm d} \rho^2}(\rho_{\rm b}(t')) \rho_{\rm b}(t')
\,G(\bm{K},t,t')
\,\Bigl(D^+(\bm{K}',t') D^+(\bm{K}-\bm{K}',t') \nonumber \\
&& \quad
      + D^+(\bm{K}',t') D^-(\bm{K}-\bm{K}',t')
      + D^-(\bm{K}',t') D^+(\bm{K}-\bm{K}',t') \nonumber \\
&& \quad
      + D^-(\bm{K}',t') D^-(\bm{K}-\bm{K}',t') \Bigr)
= (\gamma -1) F_1(\bm{K},\bm{K}',t) \,.
\label{F2(K,K',t)}
\end{eqnarray}
It is cumbersome to perform the integration
of Eqs.~(\ref{ET(K,K',t)})--(\ref{F2(K,K',t)}) in a complete form
unless $\gamma = 4/3$.
(See Ref.~\cite{moritate} for $\gamma = 4/3$.)
However, we can obtain the temporal factors in an approximate form
in the following way.
By the definition of the Bessel function,
\begin{equation}
\mathcal{J}_{\pm \nu} \left( A |\bm{K}| t^{-\gamma + 4/3}
                   \right)
= \sum_{n=0}^{\infty} \frac{(-1)^n}{n! \  \Gamma(\pm \nu + n+1)}
  \left( \frac{A |\bm{K}|}{2} \right)^{\pm \nu +2n}
  t^{\pm (5/6) + (8-6\gamma)n/3} \,,
\end{equation}
and thus if $A|\bm{K}|t^{-\gamma+4/3} \ll 1$,
we can utilize the following approximation formulae:
\begin{equation}\label{bessel-approx}
\mathcal{J}_{\pm \nu} \left( A |\bm{K}| t^{-\gamma+4/3} \right)
\simeq \frac{(A|\bm{K}|/2)^{\pm \nu}}{\Gamma (\pm \nu+1)}
\, t^{\pm 5/6} \,.
\end{equation}
Note that these formulae are useful in the case $\gamma > 4/3$,
because they give the leading term with respect to $t$
when $\gamma > 4/3$.
Substituting these formulae into Eqs.~(\ref{ET(K,K',t)})--(\ref{F2(K,K',t)}),
we have
\begin{equation}
E^{\rm T}(\bm{K}, \bm{K}', t) \simeq
\frac{A^2 (4-3\gamma)^2}{3(4-2\gamma)(13-6\gamma) \Gamma(\nu+1)^2}
\left( \frac{A^2 |\bm{K}'|
             |\bm{K}-\bm{K}'|}{4} \right)^{\nu}
t^{-2\gamma+4} \quad \mbox{for } \gamma \ne 2 \,,
\end{equation}
\begin{equation}
E^{\rm T}(\bm{K}, \bm{K}', t) \simeq
\frac{4 A^2}{3 \Gamma(-1/4)^2}
\left( \frac{A^2 |\bm{K}'|
       |\bm{K}-\bm{K}'|}{4} \right)^{-5/4}
\left( \ln t -3 \right) \quad \mbox{for } \gamma = 2 \,,
\end{equation}
\begin{equation}
E (\bm{K}, \bm{K}', t) \simeq
\frac{15 \pi}{28 \sin \nu \pi}
\frac{1}{(4-3\gamma) \Gamma (\nu+1)^3 \Gamma (-\nu+1)}
\left( \frac{A^2 |\bm{K}'|
             |\bm{K}-\bm{K}'|}{4} \right)^{\nu}
t^{4/3} \,,
\end{equation}
\begin{eqnarray}
F_1 (\bm{K}, \bm{K}', t) &\simeq&
\frac{5 \pi}{6 \sin \nu \pi}
\frac{A^2 (4-3\gamma)}
     {(5-2\gamma)(10-6\gamma) \Gamma(\nu+1)^3 \Gamma(-\nu+1)}
\left( \frac{A^2 |\bm{K}'|
             |\bm{K}-\bm{K}'|}{4} \right)^{\nu}
t^{-2\gamma+4} \nonumber \\
&& \quad \mbox{for } \gamma \ne 5/2, \  5/3 \,,
\end{eqnarray}
\begin{eqnarray}
F_1 (\bm{K}, \bm{K}', t) &\simeq&
-\frac{7 \pi}{12 \sin (5 \pi/7)}
\frac{A^2}{\Gamma(2/7)^3 \Gamma(12/7)}
\left( \frac{A^2 |\bm{K}'|
       |\bm{K}-\bm{K}'|}{4} \right)^{-5/7}
t^{-1} \left( \ln t + \frac{3}{5} \right) \nonumber \\
&& \quad \mbox{for } \gamma = 5/2 \,,
\end{eqnarray}
\begin{equation}
F_1 (\bm{K}, \bm{K}', t) \simeq
-\frac{3A^2}{80 \pi}
\left( \frac{A^2 |\bm{K}'|
       |\bm{K}-\bm{K}'|}{4} \right)^{-5/2}
t^{2/3} \left( \frac{3}{5} - \ln t \right)
\quad \mbox{for } \gamma = 5/3 \,.
\end{equation}

Let us re-examine the relation between the perturbative solutions
and the polytropic index, which seems curious in the first-order level,
as we mentioned at the end of the previous section.
In the second-order level,
the ratio of $E (\bm{K}, \bm{K}', t)$
to the other temporal factors (e.g. $F_1 (\bm{K}, \bm{K}', t)$)
can be taken as a measure of the pressure effect,
because $E (\bm{K}, \bm{K}', t)$ is
of gravitational origin and the others are of pressure origin,
and thus the ratio is similar to the Jeans wavenumber
$K_{\rm J} \propto a^{(3\gamma -4)/2}$ in the first order.
The ratio reads
\begin{equation}
\frac{E (\bm{K}, \bm{K}', t)}
     {F_1 (\bm{K}, \bm{K}', t)}
\sim A^{-2} \, t^{2\gamma - 8/3}
\sim \frac{C_1}{C_2} \, a^{3\gamma -4} \,.
\end{equation}
This means that the curious tendency of the first-order solutions
is, unfortunately, unchanged at second order, contrary to our expectation.
This result may be a consequence of the perturbation scheme we adopt.
See Sec.~\ref{sec:discuss} for detailed discussion on this point.
%

\section{Illustration in some models}\label{sec:illust}

In this section, we illustrate the perturbation theory
formulated in the previous sections
with examples in one- and two-dimensional systems.
In our previous paper~\cite{moritate}, we computed the power spectra
of density perturbations in a one-dimensional model
for the case $\gamma =4/3$.
Here we calculate the power spectra for the case $\gamma =5/3$,
and discuss the differece of the power spectra
for the variation of the polytropic index $\gamma$.
It is of significance to compute and compare the power spectra
in the Eulerian coordinates,
because the evolution of density perturbations has to be discussed
in the physical Eulerian coordinates,
and it is non-trivial how a physical variable is rewritten
due to the transformation between the Lagrangian and
the Eulerian coordinates.
Moreover, we present the density field in a two-dimensional model,
and clarify how the pressure effect appears in a spatial pattern
of the density field by comparison with the dust case.
In this section, we assume the Einstein-de Sitter background
with the scale factor $a(t) = t^{2/3}$ for simplicity.
The power spectrum of density perturbations is defined as
 $\mathcal{P}(\bm{k},t) \equiv
 \langle |\delta(\bm{k},t)|^2 \rangle$,
where $\bm{k}$ is a wave vector associated with
the Eulerian coordinates $\bm{x}$,
$\delta \equiv (\rho - \rho_{\rm b})/\rho_{\rm b}$
is the density contrast,
and $\langle \cdot \rangle$ denotes an ensemble average
over the entire distribution.

\subsection{Power spectra in a one-dimensional model}

We calculate the power spectra of density perturbations
in a one-dimensional model for the case $\gamma =5/3$.
We did this in our previous paper~\cite{moritate} for
the case $\gamma =4/3$.
Here we choose another value of $\gamma$ and see the difference
of the results for the variation of $\gamma$.
The first-order solution is then written as
\begin{equation}
\widehat{S}(K,t)
= D^+(K,t) C^+(K) + D^-(K,t) C^-(K) \,,
\end{equation}
where $K$ is a component of the direction of inhomogeneities
in the Lagrangian wave vector $\bm{K}$, and
\begin{equation}
D^{\pm}(K,t) = t^{-1/6} \mathcal{J}_{\mp 5/2}
\left( A|K|t^{-1/3} \right) \,.
\end{equation}
The Jeans wavenumber is found to be
$K_{\rm J} = \sqrt{6} t^{1/3} / A$ from Eq.~(\ref{kjeans}).

We consider how to determine $C^{\pm} (K)$ from
the initial conditions for an illustration.
Here we set the initial density contrast $\delta_{\rm in}$
and the initial peculiar velocity $\bm{v}_{\rm in}$
so that they coincide with those given by the Zel'dovich
approximation, which is the Lagrangian first-order approximation
for a dust fluid.
The Zel'dovich approximation in a one-dimensional system
is written as
\begin{equation}
x_1 = q_1 + t^{2/3} \Psi_{,1} (q_1) \,, \quad
x_2 = q_2 \,, \quad
x_3 = q_3 \,,
\end{equation}
\begin{equation}
\delta (q_1, t) = \frac{1}{1+t^{2/3} \Psi_{,11} (q_1)} -1 \,,
\end{equation}
where $\Psi(q_1)$ is an arbitrary spatial function,
describing initial inhomogeneities.
Then we have
\begin{eqnarray}
\label{delin-za}
\delta_{\rm in} &=& \frac{1}{1+ \Psi_{,11}(q_1)} -1
 \simeq -\Psi_{,11} (q_1) \,, \\
\bm{v}_{\rm in} &=& (v_{\rm in}, 0, 0)
= \left.
\left( \frac{2}{3}t^{1/3} \Psi_{,1}(q_1), 0, 0 \right)
  \right|_{t=t_{\rm in}}
= \left( \frac{2}{3} \Psi_{,1} (q_1), 0, 0 \right) \,,
\label{vin-za}
\end{eqnarray}
where we define an initial time $t_{\rm in} \equiv 1$.
As for the case $P=\kappa \rho^{5/3}$,
the first-order solution gives
\begin{eqnarray}
\label{delin-p}
\widehat{\delta_{\rm in}}(K) &=&
K^2 \left[ \mathcal{J}_{-5/2} (A|K|) C^{+}(K)
         + \mathcal{J}_{5/2} (A|K|) C^{-} (K)) \right] \,, \\
\widehat{v_{\rm in}}(K) &=& iK \left[
 \left\{-\frac{1}{6} J_{-5/2} (A|K|)
 + \frac{\rm d}{{\rm d}t} \left. J_{-5/2}(A|K|t^{-1/3})
 \right|_{t=t_{\rm in}} \right\} C^{+}(K)
 \right. \nonumber \\
&& \quad \left.
 + \left \{-\frac{1}{6} J_{5/2} (A|K|)
 + \frac{\rm d}{{\rm d}t} \left. J_{5/2} (A|K|t^{-1/3})
 \right|_{t=t_{\rm in}} \right\} C^{-} (K)
\right] \,.
\label{vin-p}
\end{eqnarray}
Comparing Eqs.~(\ref{delin-za}) and (\ref{delin-p}),
and Eqs.~(\ref{vin-za}) and (\ref{vin-p}), we find
\begin{eqnarray}
C^{+} (K) &=& -\sqrt{\frac{\pi A}{2|K|^3}}
\left( \cos(A|K|) - \frac{1}{A|K|} \sin(A|K|) \right)
\widehat{\delta_{\rm in}} (K) \,, \\
C^{-} (K) &=& -\sqrt{\frac{\pi A}{2|K|^3}}
\left( \sin(A|K|) + \frac{1}{A|K|} \cos (A|K|) \right)
\widehat{\delta_{\rm in}} (K) \,.
\end{eqnarray}
The initial density perturbation
$\widehat{\delta_{\rm in}} (K) =
|\widehat{\delta_{\rm in}} (K)| \exp (i\phi_K)$
is chosen so that
$|\widehat{\delta_{\rm in}} (K)|^2 \propto |K|^n$
with the spectral index $n=0, \pm 1$,
and the phases $\phi_K$ are randomly distributed on
the interval $[0, 2\pi]$.
We set the constant $A$ so that the Jeans wavenumber
$K_{\rm J}$ is $80$ at the initial time, $t=t_{\rm in}$,
where $a=1$.

To compute the power spectra within the Lagrangian approximations,
we have to take caution about the difference between
the Lagrangian and the Eulerian wave vectors, $\bm{K}$ and $\bm{k}$.
The Lagrangian solutions are obtained in terms of $\bm{K}$,
while the power spectra are presented by using $\bm{k}$.
Thus we have to transform the Lagrangian solutions
into the form in the Eulerian space.
The way of the transformation is described in, e.g. subsection~4.3
of Ref.~\cite{moritate}.

\begin{figure}
 \includegraphics{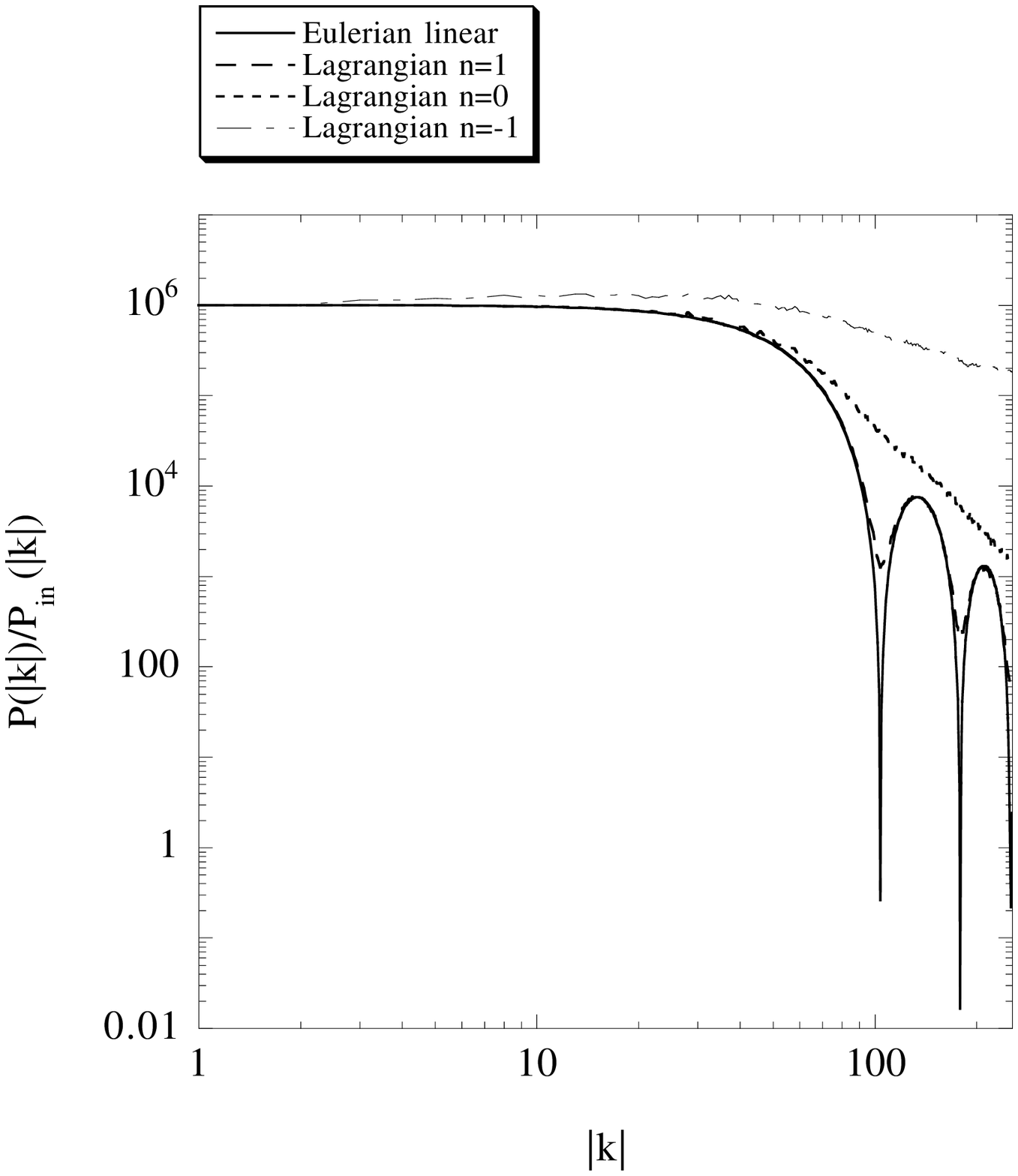}
 \caption{The `transfer function' of density perturbations
 at $a=1000$ computed by the Eulerian linear theory,
 and Lagrangian first-order approximations.
 It does not depend on the initial conditions
 in the Eulerian linear theory,
 but does in the Lagrangian approximation.}
\label{fig:1D-Trans}
\end{figure}

In Fig.~\ref{fig:1D-Trans}, we show the power spectra
$\mathcal{P}(k,t)$ at $a=1000$,
where $k$ is a component of the direction of inhomogeneities
in the Eulerian wave vector $\bm{k}$,
using the Eulerian linear theory and
the Lagrangian first-order approximation.
Instead of the power spectrum itself,
we present the `transfer function',
$\mathcal{P}(k,t)/\mathcal{P}(k,t_{\rm in})$, for convenience
because it does not depend on the initial conditions
in the Eulerian linear theory
but does in the Lagrangian approximations generally.
The spectra by the Lagrangian second-order approximation
are not presented, because they are almost coincident
with those by the first-order approximation,
as in the $\gamma=4/3$ case.
Indeed the difference between the Lagrangian first-order
and second-order approximations in the $\gamma=4/3$ case
is less than 10 \% at $|k| \alt 150$,
and that in the $\gamma =5/3$ case
becomes still smaller, less than 1 \% at $|k| \alt 150$
within our illustrations.
(See Sec.~\ref{sec:discuss} for the reason.)
%

\begin{figure}
  \includegraphics{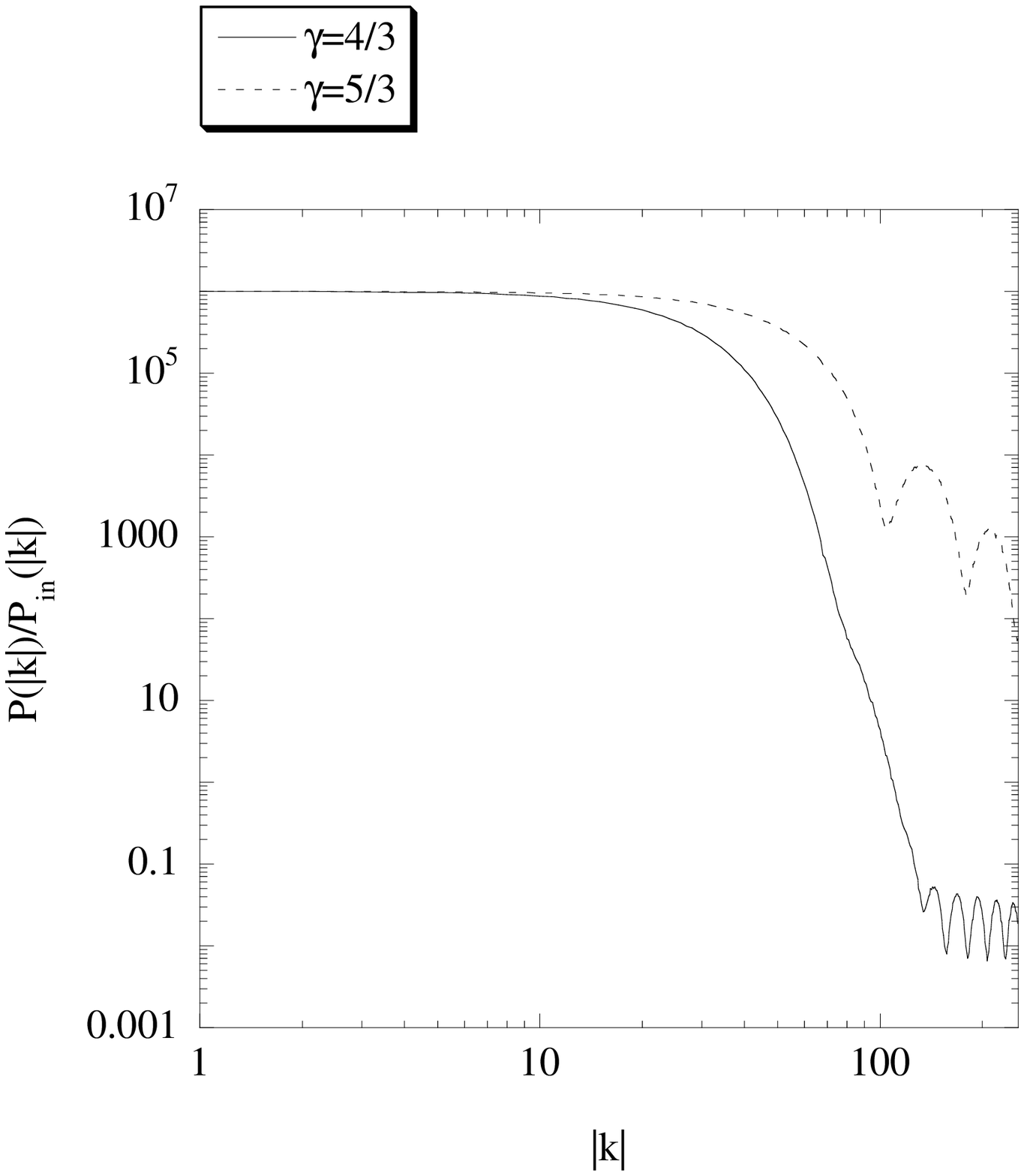}
  \caption{The `transfer function' of density perturbations
  at $a=1000$ computed by the Lagrangian first-order
  approximation in the $\gamma=4/3$ and $5/3$ cases.
  Small-scale perturbations in the $\gamma=5/3$ case
  are developed more effectively than in the $\gamma=4/3$ case.
  }
\label{fig:Trans-gamma}
\end{figure}

%
In our previous paper, we compared the Eulerian linear theory
and the Lagrangian approximations in the $\gamma=4/3$ case,
where the Jeans wavenumber $k_{\rm J}$ is a constant.
In this case, the Eulerian linear density perturbations
with wavenumbers smaller than a constant wavenumber
always grow, while those with wavenumbers larger than that
always decay with acoustic oscillation
because of the constancy of the Jeans wavenumber.
On the other hand, in the Lagrangian approximations,
small-scale perturbations are developed by the nonlinear effect,
and as a result, the difference between the Eulerian
and the Lagrangian approximations becomes large
especially at high-frequency region.
(See Fig.~2 of Ref.~\cite{moritate}.)

Now we observe the results of the $\gamma=5/3$ case,
 Fig.~\ref{fig:1D-Trans}.
In this case, the Jeans wavenumber depends on time,
and it becomes about $2500$ at $a=1000$
whereas it is set as $80$ at the initial time.
This means that the Eulerian linear density perturbations
with wavenumbers between $80$ and $2500$ are initially oscillating,
but become growing modes later.
Actually we can see this tendency at high-frequency region
in Fig.~\ref{fig:1D-Trans}.
As for the Lagrangian approximation,
small-scale perturbations are enhanced
because of the nonlinear effect, as in the $\gamma=4/3$ case.
The difference between the Eulerian and the Lagrangian
approximations is, however, not so large
because of the behavior of the Eulerian density perturbations
mentioned above.

 For comparison of the $\gamma=4/3$ and $5/3$ cases
in the Lagrangian first-order approximation,
we show in Fig.~\ref{fig:Trans-gamma} the transfer function
for both the cases, using the same initial conditions.
This figure tells us that the growth of density perturbations
computed by the Lagrangian approximation is suppressed
by the pressure more weakly in the $\gamma=5/3$ case.
This implies that the curious behavior of
the Lagrangian perturbation solutions is preserved
even if we observe it in the Eulerian coordinates.

\subsection{Density field in a two-dimensional model}

Next we consider an illustration in a two-dimensional model.
In a dust model, Buchert and Ehlers~\cite{bueh93} showed
the density field with the Zel'dovich and
the `post-Zel'dovich' approximations.
 Following their illustrations,
we present a realization of the density field
(mapped by $128^2$ particles) with our approximations,
in order to see how the pressure effect appears
in a spatial pattern.
We set the initial conditions for the scalar function
$S(\bm{q},t)$ as
\begin{eqnarray}
S(\bm{q},t_{\rm in}) = \mu \sum_{K_1} \sum_{K_2}
\frac{1}{K_1^2+K_2^2}
\left [\cos ( K_1 q_1 + K_2 q_2 + \phi(K_1, K_2)) \right] \,, \\
K_1^2+K_2^2 \neq 0, \quad
K_{1,2} = 0, 1, \cdots ,5 \,, \nonumber
\end{eqnarray}
where the phases $\phi(K_1, K_2)$ are random numbers
between $0$ and $2 \pi$,
and the amplitude $\mu$ is chosen
so that $\mu=3.0 \times 10^{-3}$.
The periodic boundary condition is imposed.
We consider the cases in which the equation of state
is given as $P=\kappa \rho^{4/3}$ and $P=\kappa \rho^{5/3}$,
assuming the Jeans wavenumber $K_{\rm J} \simeq 8$.

\begin{figure}
 \includegraphics[scale=.7]{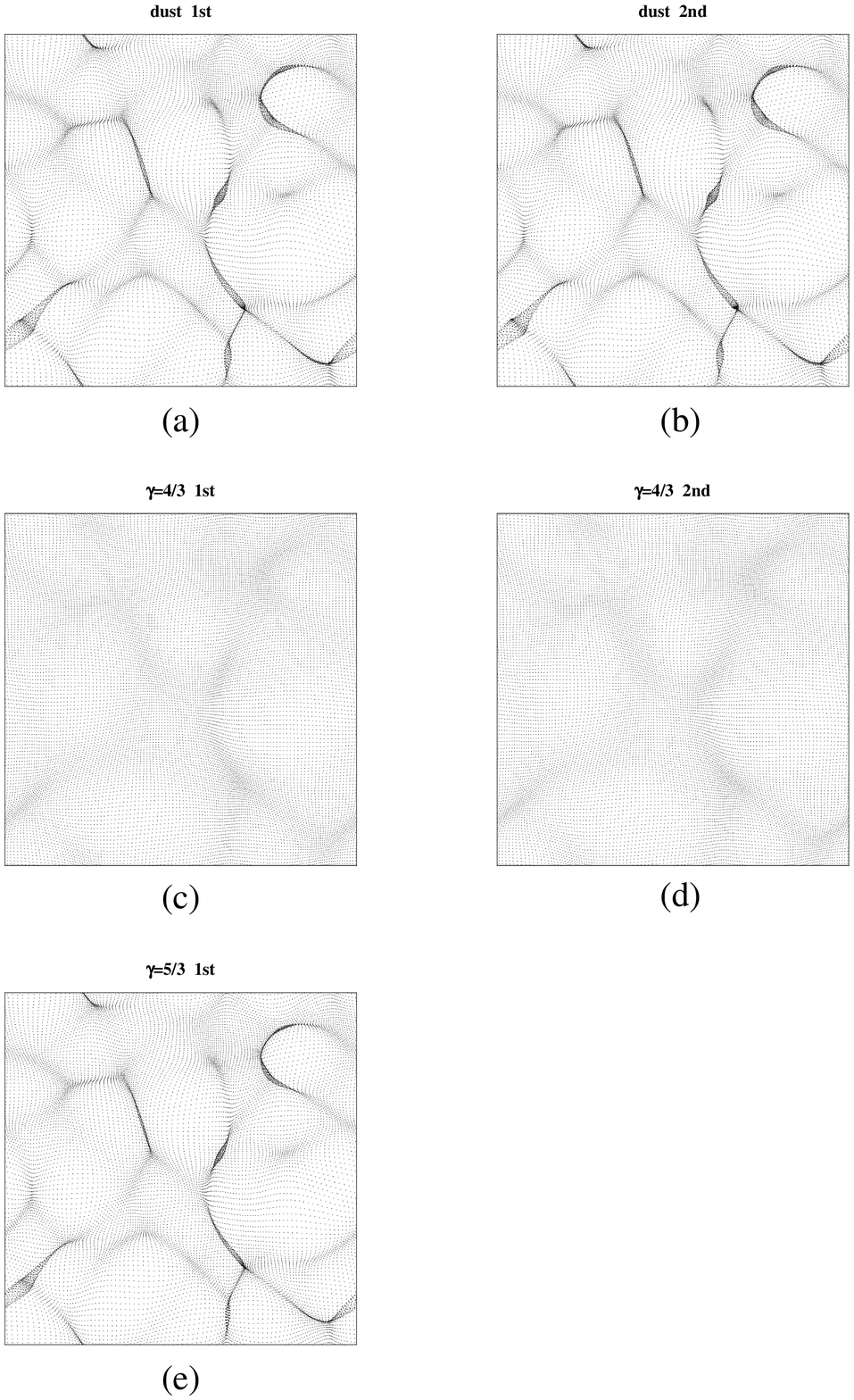}
 \caption{The particular density field of a two-dimensional model
 at $a=1000$. Shell crossings just occur in the dust case.
 (a) First-order approximation without pressure
 (the Zel'dovich approximation).
 (b) Second-order approximation without pressure
 (the `post-Zel'dovich' approximation).
 (c) First-order approximation with pressure, $\gamma =4/3$.
 (d) Second-order approximation with pressure, $\gamma =4/3$.
 (e) First-order approximation with pressure, $\gamma =5/3$.}
\label{fig:2D-a1000}
\end{figure}

\begin{figure}
 \includegraphics[scale=.7]{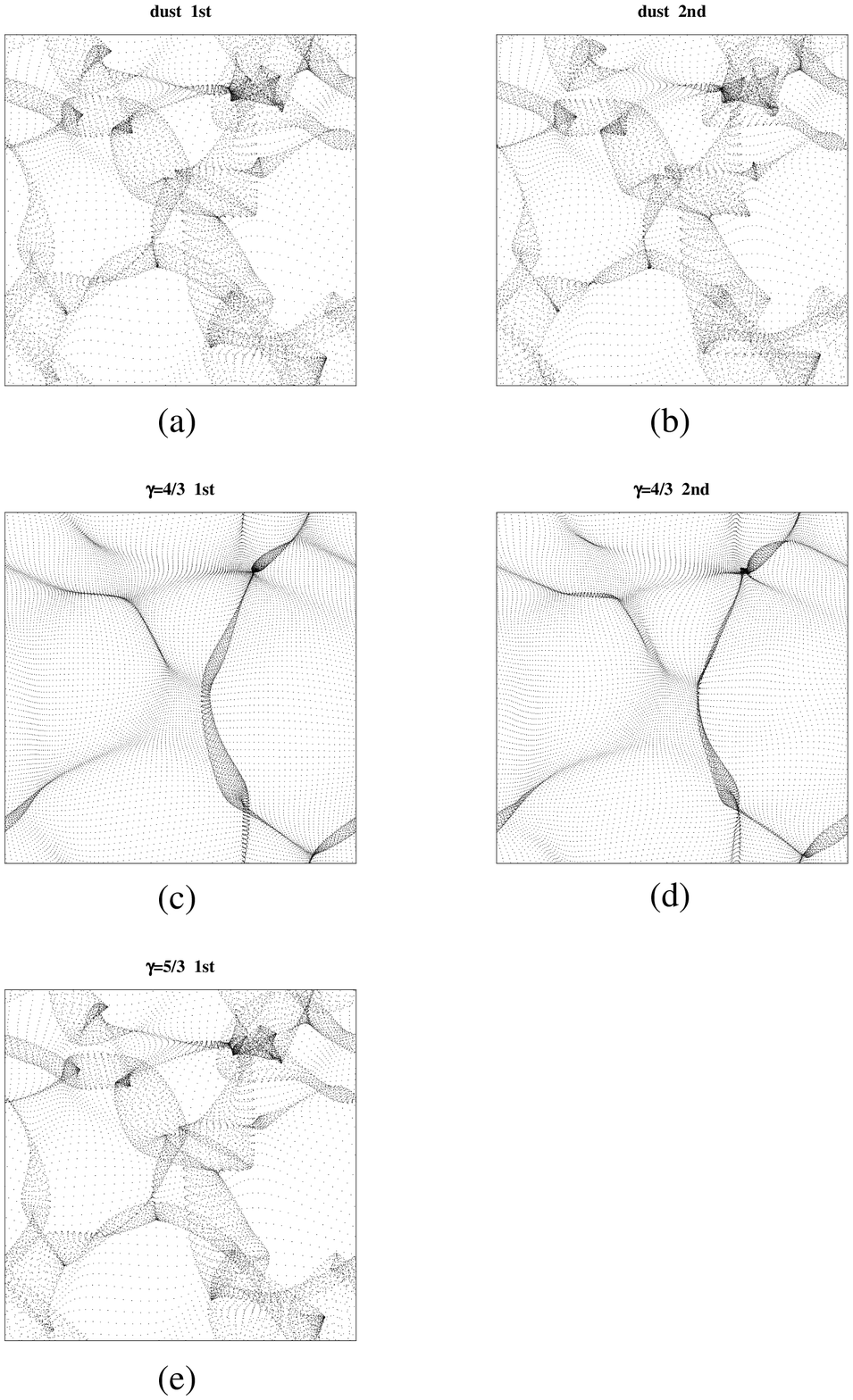}
 \caption{The particular density field of a two-dimensional model
 at $a=3000$. Shell crossings just occur in the $\gamma =4/3$ case.
 (a) First-order approximation without pressure
 (the Zel'dovich approximation).
 (b) Second-order approximation without pressure
 (the `post-Zel'dovich' approximation).
 (c) First-order approximation with pressure, $\gamma =4/3$.
 (d) Second-order approximation with pressure, $\gamma =4/3$.
 (e) First-order approximation with pressure, $\gamma =5/3$.}
\label{fig:2D-a3000}
\end{figure}

Setting the initial conditions at $a=1$,
the time evolution of the density field is shown
in Figs.~\ref{fig:2D-a1000} and \ref{fig:2D-a3000}.
In the $\gamma=4/3$ case, the evolution obviously proceeds slowly
because of the pressure effect.
In Fig.~\ref{fig:2D-a1000}, shell crossings just arise
in the dust case, while the evolution remains still
quasi-nonlinear regime in the $\gamma =4/3$ case ($|\delta| \le 1.0$).
In Fig.~\ref{fig:2D-a3000}, shell crossings are being formed
in the $\gamma =4/3$ case, while in the dust case
high-density structures are being dissolved.
In these figures, the difference between the first- and
second-order approximations seems still small
on large scales (compare (a) and (b), and (c) and (d)),
although the second-order solutions should compensate
shortcomings of the first-order approximation on small scales,
as was discussed by Buchert and Ehlers~\cite{bueh93}
for the dust case.

Above we have mentioned the $\gamma = 4/3$ case,
but what will happen if we take larger value of $\gamma$
such as $5/3$? To answer this question,
we show in Figs.~\ref{fig:2D-a1000} (e) and \ref{fig:2D-a3000} (e)
the results computed by the Lagrangian first-order approximation
in the $\gamma=5/3$ case.
(The results by the Lagrangian second-order approximation are omitted,
because we may presume them easily from other results presented.)
As we stated in Secs.~\ref{sec:first} and \ref{sec:second},
the pressure effect in this case becomes weaker than
in the $\gamma = 4/3$ case.
Indeed we see that the spatial density pattern resembles
that in the dust case, rather than that in the $\gamma =4/3$ case.

In our perturbation scheme,
shell crossings arise in the $\gamma =4/3$ and $5/3$ cases
in spite of the presence of the pressure effect,
but the features of first collapsing objects are manifestly
different from those in the dust case.
(Compare, e.g. Figs.~\ref{fig:2D-a1000} (a) and \ref{fig:2D-a3000} (c).)
The growth of small-scale structures is particularly suppressed
because of the pressure effect,
and therefore the size of overdense region becomes larger,
as if the density field was spatially coarse-grained.
(Compare, e.g. Figs.~\ref{fig:2D-a3000} (a) and (c).)
Consequently our perturbation scheme may work like
the `truncated Zel'dovich approximation'~\cite{cms,mps,ssmpm},
which yields a coarse-grained density field
of the original Zel'dovich approximation.

\section{Discussion and Concluding Remarks}
\label{sec:discuss}

We have developed a perturbation theory in the Lagrangian
hydrodynamics for a cosmological fluid with pressure.
Hydrodynamic equations in the Lagrangian coordinates
have been solved perturbatively up to second order,
extending our earlier work.
In our earlier work~\cite{moritate},
we solved the first-order perturbation equations
in the Einstein-de Sitter background,
and the second-order ones explicitly for the case $\gamma =4/3$.
In this paper, we have obtained the first-order solutions
in non-flat backgrounds and flat backgrounds with $\Lambda \ne 0$,
and the approximate second-order solutions for the case
$\gamma > 4/3$.
We have found that in several cases, the first-order solutions
are written in terms of Gauss' hypergeometric function.
We have also presented illustrations in one- and
two-dimensional systems, showing how our approximation theory
describes the evolution of cosmological inhomogeneities.

In Sec.~\ref{sec:illust}, we have computed the power spectra
of density perturbations in a one-dimensional model
for the case $\gamma = 5/3$ with the Eulerian linear theory
and the Lagrangian first-order approximation,
and have shown some amount of the difference between them.
Our numerical calculations have also shown the difference
between the Lagrangian first-order and second-order approximations,
smaller than that in the $\gamma = 4/3$ case.
Let us investigate the reason of the smallness
by considering single-wavemode perturbations
and evaluating the ratio of the second- to the first-order solution,
as we did in subsection~4.4 of Ref.~\cite{moritate}.
We assume that the first-order solution is written as
\begin{equation}
S(q_1,t) = \frac{\epsilon}{K^2}
{\rm Re} \left[
\left( c^+(K) D^+(K,t) + c^-(K) D^-(K,t) \right)
\exp(iKq_1) \right] \,,
\end{equation}
where $\epsilon$ is the amplitude of the initial
density perturbations,
$c^{\pm}(K)$ denote constants of $O(1)$,
and $D^{\pm}(K,t)$ are given by Eq.~(\ref{hatS-t}).
Then, from Eq.~(\ref{hatzeta}),
the second-order solution becomes
\begin{equation}
\zeta(q_1,t) \sim -\frac{\epsilon^2}{4\pi}
{\rm Re} \left[ F_1(2K,K,t) \exp(i2Kq_1)
            \right] \,.
\end{equation}
 For a concrete estimation, we assume $A|K|t^{-\gamma+4/3} \ll 1$
and use the approximation formulae, Eq.~(\ref{bessel-approx}),
for the cases $\gamma > 4/3$.
This assumption is reasonable because this is equivalent
to taking into account perturbation modes whose Lagrangian wavenumbers
are smaller than the Jeans wavenumber.
The first-order and second-order solutions are then reduced to
\begin{eqnarray}
S(q_1,t) &\sim& \frac{\epsilon}{K^2}
                \left( \frac{A|K|}{2} \right)^{\nu} t^{2/3}
                \, {\rm Re} [\exp(iKq_1)] \,, \\
\zeta(q_1,t) &\sim& -\epsilon^2 A^2
                \left( \frac{A|K|}{2} \right)^{2\nu}
                t^{-2\gamma +4} \, {\rm Re} [\exp(i2Kq_1)] \,,
\end{eqnarray}
where $\nu = 5/(8-6\gamma)$, and thus we find
\begin{equation}
\left| \frac{\zeta(q_1,t)}{S(q_1,t)} \right|
\alt \epsilon \left( \frac{A|K|}{2} \right)^{\nu+2}
t^{-2\gamma + 10/3}
\sim \epsilon \left( \frac{K}{K_{\rm J}} \right)^2
\left( \frac{A|K|}{2} \right)^{\nu} t^{2/3} \,.
\end{equation}
Note that the factor $\epsilon (A|K|/2)^{\nu} t^{2/3}$
corresponds to the Eulerian linear density perturbation
and is of order unity at most in our case.
Then we can show that $|\zeta/S| \ll 1$,
since the assumption $A|K|t^{-\gamma+4/3} \ll 1$
is equivalent to $|K|/K_{\rm J} \ll 1$.

In the above estimation, the second-order solution
$\zeta(q_1,t)$ is of purely pressure origin
because of the one-dimensionality,
and thus can be regarded as a measure of
the `second-order pressure effect'.
Manifestly the effect of $\zeta(q_1,t)$ becomes weaker in time
as we take the larger value of $\gamma$.
This curious fact is exactly the same as
what we have addressed at the end of Secs.~\ref{sec:first}
and \ref{sec:second}.
Now let us examine the cause of the fact.
We remark the terms of pressure origin in the perturbation
equations, Eqs.~(\ref{ddotS}), (\ref{ddotzetaL}),
and (\ref{lap-ddotzetaT}).
Then we see that all the terms of pressure origin have
time-dependent coefficients such as
${\rm d}P/{\rm d}\rho (\rho_{\rm b})$
and ${\rm d}^2 P/{\rm d}\rho^2 (\rho_{\rm b}) \rho_{\rm b}$,
which behave as
\[
\frac{{\rm d} P}{{\rm d} \rho} (\rho_{\rm b})
\propto
\frac{{\rm d}^2 P}{{\rm d}\rho^2} (\rho_{\rm b}) \rho_{\rm b}
\propto a^{-3 \gamma +3} \,,
\]
under the assumption $P \propto \rho^{\gamma}$.
These coefficients originate from the perturbation scheme,
and we can safely claim that these coefficients yield
the curious behavior of the perturbation solutions.
In addition, these coefficients will appear at any order
in the perturbation scheme,
and therefore the curious behavior will arise, i.e.
the larger value of $\gamma$ will produce the weaker effect
of pressure at any order,
as far as we consider the Lagrangian perturbation scheme.
Our two-dimensional illustration also indicates
how the evolution of inhomogeneities is sensitive
to the variation of $\gamma$; 
the pressure works effectively in the $\gamma =4/3$ case,
but does not in the $\gamma =5/3$ case,
although it depends on the choice of values of parameters in general.
Buchert et al.~\cite{bdp} argued that the $\gamma =2$ case
corresponds to the adhesion approximation~\cite{gurbatov},
but, considering our illustration, it seems difficult
to realize the adhesion-like approximation
in the $\gamma =2$ case within the Lagrangian perturbation scheme.

However, there should be no such curious matter
in the exact level of hydrodynamic equations.
To see this, let us consider the one-dimensional case,
where the relation between the Eulerian and the Lagrangian
coordinates are given as
\begin{equation}
x_1 = q_1 + s_1(q_1,t) \,, \quad
x_2 = q_2 \,, \quad x_3 = q_3 \,.
\end{equation}
Under the assumption $P=\kappa\rho^{\gamma}$,
the exact equation for $s_1$ is~\cite{adler,moritate}
\begin{equation}\label{exact1D}
\ddot{s_1} + 2\frac{\dot{a}}{a}\dot{s_1}
- 4\pi G\rho_{\rm b} s_1
- \frac{\kappa\gamma\rho_{\rm b}^{\gamma-1}}{a^2}
  \frac{s_{1,11}}{(1+s_{1,1})^{1+\gamma}} = 0 \,,
\end{equation}
where the fourth term of the left-hand side
holds the pressure effect.
This term also has the time-dependent coefficient,
${\rm d}P/{\rm d}\rho (\rho_{\rm b})$,
but simultaneously includes the effect of inhomogeneities
by $(1+s_{1,1})^{1+\gamma}$ in the denominator.
As long as $|s_{1,1}| \ll 1$,
the results of the perturbation theory are reproduced,
but once the flow lines of the fluid approach to
the shell-crossing singularities, $1+s_{1,1} \rightarrow 0$,
the effect of inhomogeneities becomes strong.
In this situation, the larger value of $\gamma$ gives
the stronger effect of pressure,
and thus no curious matter will arise.

In our earlier work, and also in this work,
we have experienced the shell-crossing problem
in spite of taking into account the pressure effect.
However, we can expect that this problem will also be avoided
in the exact level,
because the fourth term of the left-hand side of
Eq.~(\ref{exact1D}) will become very large
near shell crossing, $1+s_{1,1} \rightarrow 0$,
and will stop the growth of density enhancement.
(Some implication may be obtained by G\"{o}tz~\cite{goetz},
who solved the one-dimensional exact equation
for the case $\gamma =1$ without cosmic expansion.)

The above discussion implies that we have to admit
that our perturbation scheme yields some artificial results.
This is true, but the Lagrangian perturbation scheme is
a natural way to solve the hydrodynamic equations in cosmology,
and our formulation will give a useful tool
for large-scale structure formation in a practical sense.
It is, in principle, applicable to any cosmological situation
in which velocity dispersion arises and
is written as a function of the density only.
Actually Fig.~\ref{fig:2D-a3000} has shown that
our scheme works better than the Zel'dovich approximation
beyond shell crossing,
giving some kind of spatial coarse graining
of the density field, as is given by
the truncated Zel'dovich approximation~\cite{cms,mps,ssmpm}.
Detailed analyses of comparison of our scheme
and the truncated Zel'dovich approximation
(and also the adhesion approximation) will be provided
in a separate publication.

As for the shell-crossing problem,
Matarrese and Mohayaee~\cite{mamo} have treated it
in Lagrangian perturbative approach for two-component fluid.
They also experienced shell crossing
in usual perturbative Lagrangian approach,
and introduced the `stochastic adhesion' model
to overcome the problem.
It will be interesting to probe how to treat the dynamics
when shell crossing is occurring,
or how to avoid shell crossing by taking account of
the pressure effect in a sophisticated manner.

\begin{acknowledgments}
We would like to thank Yasuhide Sota for useful discussion
in the early stage of the work.
MM also thanks Thomas Buchert for the hospitality
during his stay in Munich,
where the final part of the work was done.
\end{acknowledgments}


\end{document}